\documentclass[letterpaper, 10 pt, conference]{ieeeconf}  % Comment this line out if you need a4paper
\IEEEoverridecommandlockouts
\usepackage{graphics} % for pdf, bitmapped graphics files
\usepackage{epsfig} % for postscript graphics files
\usepackage{booktabs} % table rules
\usepackage{hyperref}
\usepackage{multirow}
\usepackage{amsmath,amssymb,amsfonts}
\usepackage{todonotes}
\usepackage{comment}
\usepackage[ruled, vlined]{algorithm2e}
\usepackage{graphicx}
\usepackage{subfigure}

\usepackage{makecell}

\usepackage{xcolor}
\usepackage[load-configurations=version-1]{siunitx}

\title{\LARGE \bf
Exploiting Multiple EEG Data Domains with Adversarial Learning
}

\author{ 
David Bethge$^{1,2}$, Philipp Hallgarten$^{1,3}$, Ozan Özdenizci$^{4,5}$, \\ 
Ralf Mikut$^{3}$, Albrecht Schmidt$^{2}$, Tobias Grosse-Puppendahl$^{1}$
\thanks{$^{1}$ Dr. Ing. h.c. F. Porsche AG, Stuttgart, Germany.}%
\thanks{$^{2}$ Ludwig-Maximilians University, Munich, Germany.}% 
\thanks{$^{3}$ Karlsruhe Institute of Technology, Karlsruhe, Germany.}%
\thanks{$^{4}$ Institute of Theoretical Computer Science, TU Graz, Austria.}%
\thanks{$^{5}$ TU Graz-SAL DES Lab, Silicon Austria Labs, Graz, Austria.}%
\thanks{Corresponding author: P.~Hallgarten (philipp.hallgarten1@porsche.de).}
}

\begin{document}

\maketitle

\begin{abstract}
Electroencephalography (EEG) is shown to be a valuable data source for evaluating subjects' mental states.
However, the interpretation of multi-modal EEG signals is challenging, as they suffer from poor signal-to-noise-ratio, are highly subject-dependent, and are bound to the equipment and experimental setup used, (\textit{i.e. domain}).
This leads to machine learning models often suffer from poor generalization ability, where they perform significantly worse on real-world data than on the exploited training data.
Recent research heavily focuses on cross-subject and cross-session transfer learning frameworks to reduce domain calibration efforts for EEG signals.
We argue that multi-source learning via learning domain-invariant representations from multiple data-sources is a viable alternative, as the available data from different EEG data-source domains (e.g., subjects, sessions, experimental setups) grow massively.
We propose an adversarial inference approach to learn data-source invariant representations in this context, enabling multi-source learning for EEG-based brain-computer interfaces.
We unify EEG recordings from different source domains (i.e., emotion recognition datasets SEED, SEED-IV, DEAP, DREAMER), and demonstrate the feasibility of our invariant representation learning approach in suppressing data-source-relevant information leakage by $35\%$ while still achieving stable EEG-based emotion classification performance.
\end{abstract}


\section{Introduction}
\label{sec:intro}

Electroencephalogram (EEG) based brain-computer interface (BCI) systems aim to identify users' intentions from brain recordings with potential uses in neurorehabilitation systems \cite{machado2010eeg}.
However, moderate decoding accuracies have limited the practical use of BCIs~\cite{xu2020cross,ozdenizci2019information}.
Due to the high data collection efforts and costs, EEG datasets highly diverge in their recording environment (\textit{e.g.}, stimulus), the equipment and devices, and the ground truths derived.
Shortage of large and homogeneous BCI datasets limits the choice of applicable models and causes a high effort if individual models are to be used for each domain.
Imbalance of EEG data source domains for classification is therefore prevalent and posing important challenges for EEG-based BCIs.

Transfer learning across different data domains as such has been extensively studied over the past decades in computer vision \cite{tzeng2017adversarial,zhuang2020comprehensive}, proposing convolutional neural networks (CNNs) to extract domain-invariant features for image search and classification across domains.
Subsequently, transfer learning on neurophysiological recording datasets (\textit{e.g.,} EEG) is becoming an active research field \cite{wan2021review}. 
Generalized neural decoder learning for across recording modalities (multi-corpus) on electrocorticography data has been recently proposed by \cite{peterson2021generalized}. 
Their approach was shown to generalize to new participants and recording modalities, robustly handle variations in electrode placement, and allow participant-specific fine-tuning with minimal data.
Also recently, \cite{xu2020cross} discussed an online pre-alignment strategy for aligning the motor imagery EEG recording distributions of different subjects before training and inference processes, and showed to significantly improve generalization across datasets.
Towards a similar goal, \cite{ozdenizci2020learning,ozdenizci2019biometrics} proposed an invariant representation learning scheme using adversarial inference to enable cross-nuisance transfer learning in EEG signal classification with deep neural networks. 
Empirical assessments on EEG decoding tasks revealed that cross-subject~\cite{ozdenizci2020learning} or cross-session~\cite{ozdenizci2019biometrics} representations can be learned with such models.
Cross-subject EEG transfer learning have been also explored for emotion recognition to generalize existing models to new subjects, and thereby reducing the demand for the calibration data amount effectively for new subjects~\cite{li2019multisource}.

In light of recent work on enabling multi-corpus learning from neurophysiological data~\cite{xu2020cross,ross2020unsupervised,rodrigues2020dimensionality, bethge2022domain}, we propose an adversarial machine learning approach to unify different raw EEG time-series and pre-process them accordingly.
Unlike previous work that has focused on learning scenarios across subjects or sessions, we explore dataset-invariant representations via an adversarial learning framework that can be used in EEG multi-label settings.
Our approach aims at expressing robust task-relevant EEG features in a latent representation for emotion recognition across several datasets, by limiting the representation to not learn nuisances specific to these datasets, hence being dataset invariant.
We evaluate our framework against the competing baseline of a state-of-the-art deep learning encoder-classifier network trained on the unified set of all data sources.

Our contributions in this work are as follows: (1) We present a unifying EEG pre-processing framework for fusing different raw EEG time-series datasets and associated emotion state labels for transfer learning. (2) We propose an adversarial machine learning framework on multivariate EEG time-series to learn dataset-invariant representations to predict EEG class labels. (3) We present an experimental study on assembling four publicly-available EEG datasets in the field of emotion recognition, and show that our approach can learn dataset-invariant representations \textit{i.e.}, transfer emotion-relevant EEG signals across datasets containing data from different subjects and measurement conditions.

\section{Methods}
\label{sec:methods}

\subsection{Notation and Problem Statement}
\label{subsec:notation}

Let $\{(X_i, y_i, s_i, d_i)\}_{i=1}^n$ denote the data samples consisting observations from a data generation process with $X \sim p(X|y,s,d), y \sim p(y)$, $s \sim p(s)$, and $d \sim p(d)$, where $X_i \in \mathbb{R}^{C_i \times T_i}$ is the raw trial EEG data from data-source $d_i$ during trial $i$ recorded from $C_i$ channels for $T_i$ discretized time samples, $y_i$ is the corresponding emotion label, that can either be a discrete state $y_i \in \{1,..,Y\}$ or a vector $y_i \in \mathbb{R}^{Y}$ depending on the data-source, and $s_i \in \{1, 2, \ldots, S\}$ denotes the subject identification (ID) number for the participant that the trial EEG data is recorded from. Since the subject IDs and emotional labels are defined and used differently within different data-sources, it is necessary to pre-process the data, as described in more detail in Sec.~\ref{sub:emotion-label-conversion}.
To describe the data-source origin of a particular EEG epoch, $d_i \in \{1,\ldots,D\}$ specifies the data-source ID of $X_i$. Note that for our problem of interest, the underlying assumption here is $s$ and $y$ as well as $d$ and $y$ being marginally independent, \textit{i.e.} the probability of a certain emotion is the same for all subjects and across all data-sources. We achieve this by balancing the samples with respect to the subject IDs and the data-source IDs, as further described in Sec.~\ref{sub:balancing-the-samples}.

We can distinguish two approaches to combine multiple data-sources in a learning pipeline: (1) pre-processing the samples and labels of the data-sources, so that they can be processed by the same encoder framework, and (2) training individual encoder frameworks for each data-source, while ensuring a consistent latent representation among all frameworks. For the scope of this paper, we will investigate the first approach, whereas our adversarial training pipeline is applicable to both.
Given training data the aim is to learn a discriminative model that predicts $y$ from observations $X$. For such a model to be generalizable across  datasets, ideally, the predictions should be invariant to $d$, which will be unknown at test time. Herein, we regard $d$ as nuisance parameters involved in the EEG data generation process and aim to learn a parametric model that can be generalized across different data sources and learns robust representations.% A similar invariant representation learning methodology was proposed in previous work \cite{ozdenizci2020learning}.

% \subsection{Approaches to Dataset Normalization and Learning}

% In general, we view two approaches to combine multiple datasets in a learning approach: 
% 1. Making the datasets as similar as possible in a preprocessing step will result in discarding information due to different ground truth labels, EEG device-dependent number of channels, and sampling frequencies. 
% 2. Training multiple singular encoders and decoders to account for variations in labels, channels, and sampling frequencies while ensuring a consistent latent space among all encoders. 

% For the scope of this paper, we will investigate approach 1.... Why a shared encoder approach?  

\subsection{Adversarially Learned Invariant EEG Representations}

\begin{figure}[t!]
    \centering
    \includegraphics[width=1\linewidth]{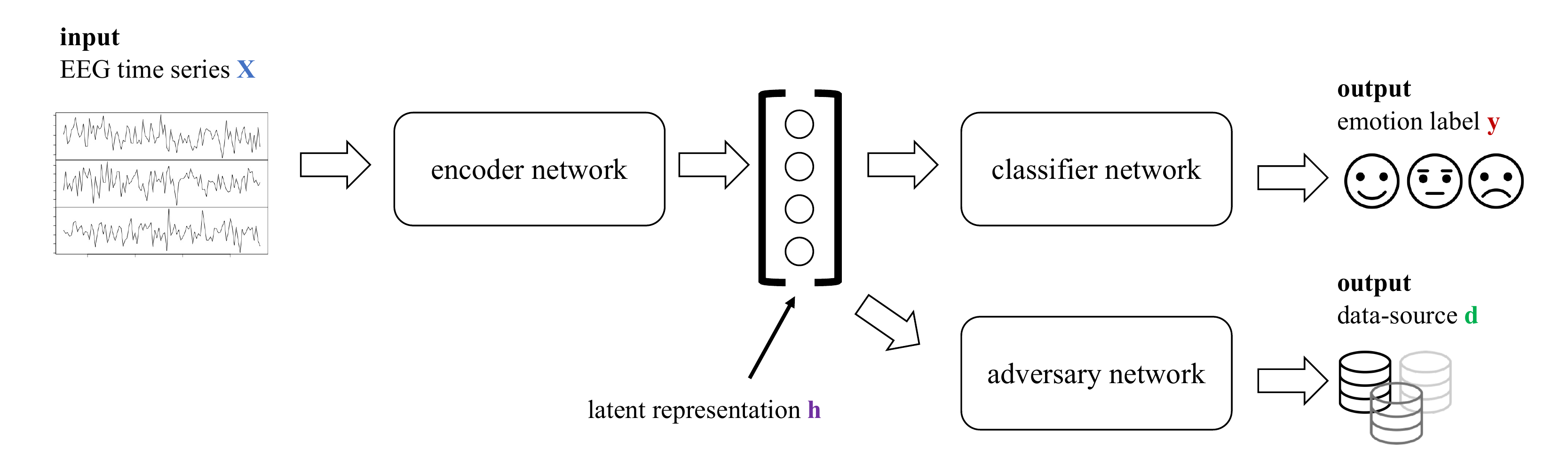}
    \caption{Overview of the network architecture proposed for adversarial domain adaptation across multiple EEG datasets, consisting of an encoder and two separate dense layer classifiers (\textit{i.e.}, a classifier network for emotions and an adversary network that identifies the EEG data-source ID).}
    \label{fig:architecture}
\end{figure}

We train a deterministic encoder with parameters $\theta_{enc}$ to learn representations $h = f(X ; \theta_{enc} )$ given the training data. We discuss the encoder network specifications in detail in Sec.~\ref{subsec:network_architecture}. 
Obtained representations $h$ are used as input separately to both a classifier network with parameters $\theta_{clf}$ to estimate $y$, as well as an adversary network with parameters $\theta_{adv}$, which aims to recover the data-source variable $d$.
Respectively, the classifier and adversary networks estimate the likelihoods $q_{{clf}} (y|h)$ and $q_{{adv}} (d|h)$.

We aim to filter factors of variation caused by $d$ within $h$. To achieve this, we propose an adversarial learning scheme. The adversary network is trained to predict $d$ by maximizing the likelihood $q_{adv} (d|h)$. At the same time, the encoder is trying to conceal information regarding $d$ that is embedded in $h$ by minimizing that likelihood, as well as retaining sufficient discriminative information for the classifier to estimate $y$ by maximizing $q_{clf} (y|h)$. Overall, we simultaneously train these networks towards the following objective:
\begin{equation}
    \label{eq:theta_optimization}
    \min_{\theta_{enc},\theta_{clf}} \max_{\theta_{adv}}\,\mathbb{E}[-\log  q_{\theta_{clf}} (y|h) + \lambda \log q_{\theta_{adv}} (d|h)]
\end{equation}
with $\theta_{enc}$ represented through $h = f (X ; \theta_{enc})$. A higher adversarial regularization weight $\lambda>0$ enforces stronger invariance from $d$ trading-off with discriminative performance. We use stochastic gradient descent (or ascent) alternatingly for the adversary and the encoder-classifier networks to optimize Eq.~\eqref{eq:theta_optimization}. %This approach is motivated by the work on adversarially learned invariant EEG representations for discriminative model training \cite{ozdenizci2020learning}. 
Note that setting the regularization parameter $\lambda=0$ indicates training a regular neural network.

\begin{algorithm}[tb]
\SetKwInOut{Input}{input}\SetKwInOut{Output}{output}
\SetAlgoLined
\Input{$\eta$ learning rate, $\lambda$ adversarial reg. weight}
\For{$epoch\leftarrow 0$ \KwTo $epochs$}{
    \For{$batch\leftarrow 0$ \KwTo $batches$}{
    \emph{\# Forward pass the input through the encoder to compute representation}\\
        $h \leftarrow f(X; \theta_{enc} )$ \\
        \emph{\# Update parameters of the adversary}\\% using $h_i$ without updating the encoder or classifier}\\
        %$\theta_{adv} \leftarrow \theta_{adv} \lambda\alpha\nabla_{\theta_{adv}}\mathcal{L}(h_i,d_i)$\\
        $\theta_{adv} \leftarrow \theta_{adv} - \eta\nabla_{\theta_{adv}} \displaystyle\mathbb{E}[-\lambda \log q_{\theta_{adv}}(d|h)]$ \\
        \emph{\# Update parameters of the encoder and emotion classifier}\\% without updating the adversary}\\
        $\theta_{enc} \leftarrow \theta_{enc} - \eta\nabla_{\theta_{enc}} \displaystyle\mathbb{E}[-\log  q_{\theta_{clf}} (y|h) + \lambda \log q_{\theta_{adv}} (d|h)]$\\
        $\theta_{clf} \leftarrow \theta_{clf} - \eta\nabla_{\theta_{clf}} \displaystyle\mathbb{E}[-\log  q_{\theta_{clf}} (y|h) + \lambda \log q_{\theta_{adv}} (d|h)]$
        %$\theta_{enc} \leftarrow \theta_{enc} + \alpha\nabla_{\theta_{enc}}\mathcal{L}(h_i,y_i,d_i)$\\
        %$\theta_{clf} \leftarrow \theta_{clf} + \alpha\nabla_{\theta_{clf}}\mathcal{L}(h_i,y_i)$\\
    }
 }
 \caption{Adversarial multi-source EEG neural network training scheme.}
 \label{algo:training}
\end{algorithm}

\subsection{Neural Network Architecture and Training}
\label{subsec:network_architecture}

Proposed model is illustrated in \autoref{fig:architecture}. 
The encoder network maps each input sample $X_i$ to a latent representation vector $h_i$, which is used as input to two separate single dense layer classifiers.
The first classifier, \textit{i.e.} \textit{emotion classifier}, predicts an EEG class label $y_i$, \textit{i.e.,} an emotional label.
The second classifier, \textit{i.e.} \textit{adversary network}, serves as an EEG domain classifier and predicts the data-source ID $d_i$ of the current training data sample. 
To solve the objective in Eq.~\ref{eq:theta_optimization}, we update the parameters of the adversary network (i.e., domain classifier) and the encoder-emotion classifier network pair alternatingly on each batch. 
Our model training pipeline is outlined in Algorithm~\ref{algo:training}. 
While the proposed architecture is not restrictive to any neural network specification, during our evaluations, for the encoder we used the state-of-the-art convolutional DeepConvNet EEG encoder backbone~\cite{schirrmeister2017deep}.

\begin{table*}[ht!]
\begin{center}
\centering
\caption{Details on the used four dataset specifications.}
\resizebox{1\textwidth}{!}{
\begin{tabular}{c l l l l l}
        \toprule
        & & \textbf{SEED}~\cite{seed} & \textbf{SEED-IV}~\cite{seed_iv} & \textbf{DEAP}~\cite{DEAP-Dataset} & \textbf{DREAMER}~\cite{katsigiannis2017dreamer}  \\
        \midrule
        \multirow{5}{*}{\rotatebox[origin=c]{90}{\textbf{Exp. Setup}}} &
        subjects (male / female) & 15 (7/8) & 15 (7/8) & 32 (16/16) & 23 (14/9) \\
        & sessions per subject & 3 & 3 & 1 & 1 \\
        & trials per session & 15 & 24 & 40 & 18 \\
        & stimuli & film clips & film clips & music videos & film clips or music videos \\
        & provided stimuli duration & $\sim$\SI{4}{\min} & $\sim$\SI{2}{\min} & 1 min & $\sim$\SIrange{1}{7}{\min} \\
        \midrule
        \multirow{4}{*}{\rotatebox[origin=c]{90}{\textbf{EEG}}} &
        number of channels & 62 & 62 & 32 & 14 \\
        & sampling rate & \SI{200}{\hertz} & \SI{200}{\hertz} & \SI{128}{\hertz} & \SI{128}{\hertz} \\
        & freq. filtering & \SIrange{0}{75}{\hertz} & \SIrange{1}{75}{\hertz} & \SIrange{4}{45}{\hertz} & - \\
        & baseline removal & - & - & 3s & - \\
        \midrule
        \multirow{5}{*}{\rotatebox[origin=c]{90}{\textbf{Labels}}} &
        emotion representation model & discrete & discrete & dimensional & dimensional \\
        & self-analysis & No & No & Yes & Yes \\
        & discrete or continuous & discrete  & discrete & continuous (1--9) & discrete (1--5) \\
        & states / dimensions & \textit{Negative, Neutral, Positive} & \textit{Sad, Fear, Neutral, Happy} & \makecell{\textit{Valence, Arousal, Dominance,} \\ \textit{Liking, Familiarity}} & \textit{Valence, Arousal, Dominance} \\
        \bottomrule
    \end{tabular}}
    \label{tab:overview-datasets}
    \end{center}\vspace{-.1cm}
\end{table*}

\section{Experimental Study Design}
\label{sec:expstudy}

\subsection{Datasets}
We regard four commonly-used and open-source EEG-based emotion recognition experiment datasets as our data-sources, namely SEED \cite{seed}, SEED-IV \cite{seed_iv}, DEAP \cite{DEAP-Dataset}, and DREAMER \cite{katsigiannis2017dreamer}. All datasets contain EEG signals recorded from multiple subjects that were exposed to audio-visual stimuli such as music videos. The EEG signals are labeled with the emotion, that the subject is assumed to have felt during recordings.
The datasets mainly differ from each other in three points. First, the experimental setup used for recording, including the number of sessions performed per subject or the used emotional stimuli. Second, the characteristics of the EEG signals, including the number of electrodes (channels) used or the sampling rate. And third, the emotion representation model used to determine the ground truth for the signals.
An overview over the specifications of the used datasets can be found in \autoref{tab:overview-datasets}.

\subsection{EEG Pre-Processing}

% \begin{table}[h!]
% \begin{center}
% \centering
% \caption{}
% \label{tab:my-table}
% \resizebox{0.5\textwidth}{!}{%
% \begin{tabular}{@{}lllll@{}}
% \hline
% dataset & channels & sampling rate (Hz) & device      & label space                              \\ \hline
% SEED    & 62       & 1000               & ESI Neuroscan  &discrete (happy, sad, neutral)           \\
% SEED-IV & 62       & 1000               & ESI Neuroscan  & discrete (happy, sad, neutral, fear)     \\
% DREAMER & 14       & 128                & Emotiv EPOC   & continuous (valence, arousal, dominance) \\
% DEAP & 32      & 512                & Emotiv EPOC   & continuous (valence, arousal, dominance) \\
% \hline
% \end{tabular}%
% }
% \end{center}
% \end{table}
We use only channels $C_i$ which are within all datasets, \textit{i.e.,} $C =$ \{'AF3', 'AF4', 'F3', 'F4', 'F7', 'F8', 'FC5', 'FC6', 'O1', 'O2', 'P7', 'P8', 'T7', 'T8'\}.
%We use only channels $C_i$ which are within all datasets\footnote{\textit{i.e.,} $C =$ \{'AF3', 'AF4', 'F3', 'F4', 'F7', 'F8', 'FC5', 'FC6', 'O1', 'O2', 'P7', 'P8', 'T7', 'T8'\}}. % (see Eq. \ref{eq:channel_subsampling}).
%  \begin{align}
%     C &= \bigcap_{i=1}^D C_i\label{eq:channel_subsampling} \\
%     f &= \min_{i=1}^D \{ f_i \} \label{eq:f_subsampling}
% \end{align}
Furthermore, we downsample all recordings to the minimum sampling rate of the datasets, \textit{i.e.,} \SI{128}{\hertz}. %(see Eq. \ref{eq:f_subsampling}).
This downsampling procedure ensures that the model can analyze the EEG time-frequency patterns coherently with the same encoder architecture.
The non-zero averages of some of the EEG Signals would lead to increased activation within the neural network. Therefore we calculate the mean value of each channel during the first three seconds of each experiment and subtract it from the whole time series. As some of the EEG signals provided to us are already bandpass filtered using different cut-off frequencies, we bandpass-filter the signals again, using a Butterworth bandpass filter, preserving the smallest common frequency-band all examples contain, \textit{i.e.,} \SIrange{4}{45}{\hertz}.

Finally, all the time series are cut into 2 seconds non-overlapping windows, resulting in a data sample space of dimension $\mathbb{R}^{n \times 14 \times 256}$ where $n$ is the number of window segments~\cite{candra2015investigation}. Note that in doing so, we make a rather weak assumption that the emotion representation in EEG is stable throughout the experiment, which makes the problem harder for us with the presence of noisy labels.

Overall, we note that through the downsampling and channel selection (least common divisor approach), we discard valuable (high-frequency) EEG information, which poses a limitation of our model's classification performance.

\subsection{Emotion Category Label Conversion}
\label{sub:emotion-label-conversion}

To date, no unified emotion model across datasets exists, and the various established models can often only be partially compared or mapped into one another. 
Among the datasets we used, two (SEED and SEED IV) employ a discrete state emotion model.
%In SEED, the three emotional states \emph{positive, neutral and negative} are distinguished, in SEED IV, the four emotional states \emph{happy, sad, neutral}, and \emph{fear}.
In contrast, DEAP and DREAMER used a dimensional model by assessing each emotion by a quantitative expression in several dimensions. The three established dimensions \emph{Valence, Arousal}, and \emph{Dominance} are used in both DEAP and DREAMER.
%DEAP additionally extends the model by the dimensions \emph{familiarity} and \emph{liking}. DEAP rates the emotions on a continuous scale from 1 to 9 in these dimensions, while DREAMER rates them on a discretized scale from 1 to 5 using a step size of 1.

For our experimental studies, we transformed the different emotion representations into a common representation. 
Since the discrete states are not differentiated enough to be reasonably mapped into a dimensional model, we converted all representations into a discrete emotion model with three states (\emph{negative, neutral, positive}). 
As the SEED dataset already uses these states, no transformation was necessary. 
By assuming that in the dimensional emotion models of DEAP and DREAMER there are four clusters associated with the states \emph{sad}, \emph{fear}, \emph{happy}, and \emph{neutral} we first transferred these representations into a discrete emotion representation model using k-means clustering.
To map the states of SEED-IV to our emotion label representation, we made the rather reasonable assumption that \emph{fear} and \emph{sad} are negative emotions and \emph{happy} is a positive emotion. 
Merging the two negative states, we were then able to transform the label representation into the required emotion state model.

\subsection{Balancing the Samples Across Data Sources}
\label{sub:balancing-the-samples}
As described in Sec.~\ref{subsec:notation}, we assume $y_i$ and $s_i$ as well as $y_i$ and $d_i$ to be marginally independent. To obtain the same distributions $p(y|s_i)$ for all subjects $s_i$ and $p(y|d_i)$ for all data-source IDs $d_i$, we balanced the samples $X_i$ with respect to the emotion label first for all subjects individually and later for all data-sources individually. We also took the same number of subjects with the same data-source IDs, giving us a fully-balanced dataset. 
Using a fully balanced and stratified dataset as such allows us to eliminate biased predictions due to imbalanced samples and ensures that our approach is enforced to not biased on certain participants, data-sources or emotion class labels. %Even though it ended up limiting us to use only a small share of the different data-sources possible, we adhered to this methodology in order to test our preliminary hypotheses in the current pilot work.
%However, to test our preliminary hypotheses in the current pilot work, we adhered to this methodology.

\subsection{Experimental Configurations}

We evaluated our model using (1) pre-processed EEG time series in conjunction with the deep neural network (DNN) architecture, and (2) manually extracted power spectral density (PSD) features from the preprocessed time series as input \cite{seed_iv}.
%PSD features were calculated within the delta (\SIrange{1}{4}{\hertz}), theta (\SIrange{4}{7}{\hertz}), alpha (\SIrange{8}{13}{\hertz}), beta (\SIrange{13}{30}{\hertz}), and gamma ($>$ \SI{30}{\hertz}) band for each sample and channel individually. The 5 values thus obtained for each of the 14 channels were then combined into one single vector. A multilayer perceptron (MLP) encoder was used for mapping the $n \times 70$-dimensional extracted PSD features to the latent representation $h$.
\footnote{PSD features were calculated within the delta (\SIrange{1}{4}{\hertz}), theta (\SIrange{4}{7}{\hertz}), alpha (\SIrange{8}{13}{\hertz}), beta (\SIrange{13}{30}{\hertz}), and gamma ($>$ \SI{30}{\hertz}) band for each sample and channel individually.
%The 5 values for each of the 14 channels were then combined into one single vector. A multilayer perceptron (MLP) encoder was used for mapping the $n \times 70$-dimensional extracted PSD features to the latent representation $h$.
}
In order to also test binary classification performance, in a different set of experiments we omitted the observations with a neutral emotion label and evaluated binary classification using the same time-series DNN architecture.

We performed five repetitions of each experiment by using $60\%$ of the preprocessed dataset as the training set, $20\%$ as the validation set, and $20\%$ as the test set. 
We ensured that the specified requirements from Sec.~\ref{sub:balancing-the-samples} was held for each of these sets.
Maximum number of epochs was always set to 500 with validation loss based early stopping (which generally resulted in completion around 50 epochs).

\section{Experimental Results}
\label{sec:results}

\subsection{Investigating Domain-Specific Leakage during Training}

For preliminary verification purposes, we monitored the dataset domain specific information leakage throughout training. We assess this by observing (1) the predictions made by the adversary network throughout epochs, as well as (2) an independent naïve Bayes classifier that is fitted per epoch on the current latent representation to predict the dataset ID.

\begin{figure}[h!]
\centering
\subfigure[]{\includegraphics[width=.97\columnwidth]{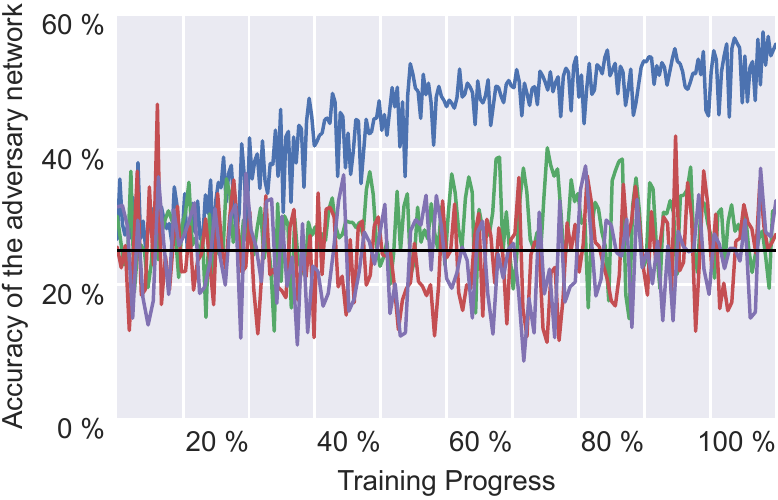}\label{fig:leakageadv}}
\subfigure[]{\includegraphics[width=.97\columnwidth]{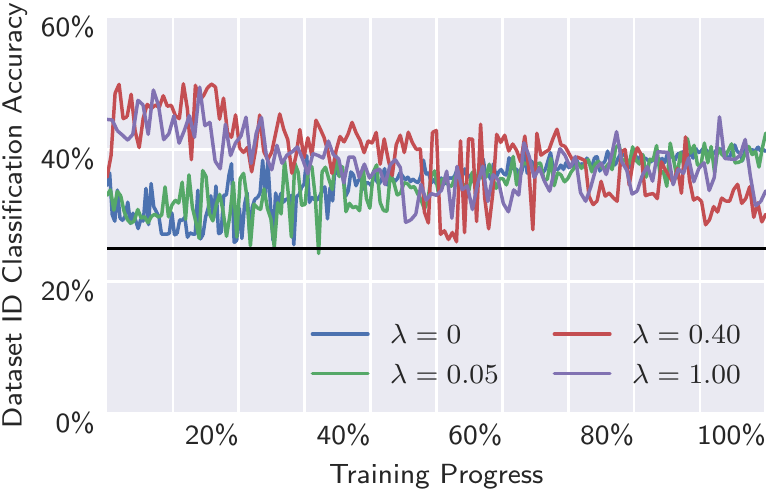}\label{fig:leakagenb}}
\caption{Domain-relevant leakage throughout training by (a) observing accuracy of the adversary network, (b) fitting Naïve Bayes classifiers to predict $d$ from $h$, for different adversarial censoring hyperparameter choices $\lambda$. The training progress is normalized to percentage by the early stopping end epoch. The black line indicates the chance level.}\vspace{-.1cm}
\label{fig:leakage-monitoring}
\end{figure}

Figure~\ref{fig:leakageadv} illustrates the prediction accuracies of the adversary network during training. Note that for the baseline model with $\lambda=0$, an adversary was still trained alongside the classifier to simply monitor $d_i$-relevant information leakage, without impacting the total loss or gradient-based parameters updates of parameters. We observe that adversarially censored models yield chance-level dataset prediction accuracies, whereas the baseline models show undesired dataset-relevant information leakage throughout training.

We present the results of the independently epoch-wise fitted naïve Bayes classifier in Figure~\ref{fig:leakagenb}. We observe that for higher $\lambda$ values (hence imposing stronger domain-invariance) estimated leakage starts to decrease with trained epochs, which again implies that our approach leads the encoder to reduce the $d$-relevant leakage in the latent space.

\subsection{Impact of Adversarial Learning on Classification}

\begin{figure}[t!]
\centering
\includegraphics[width=.97\columnwidth]{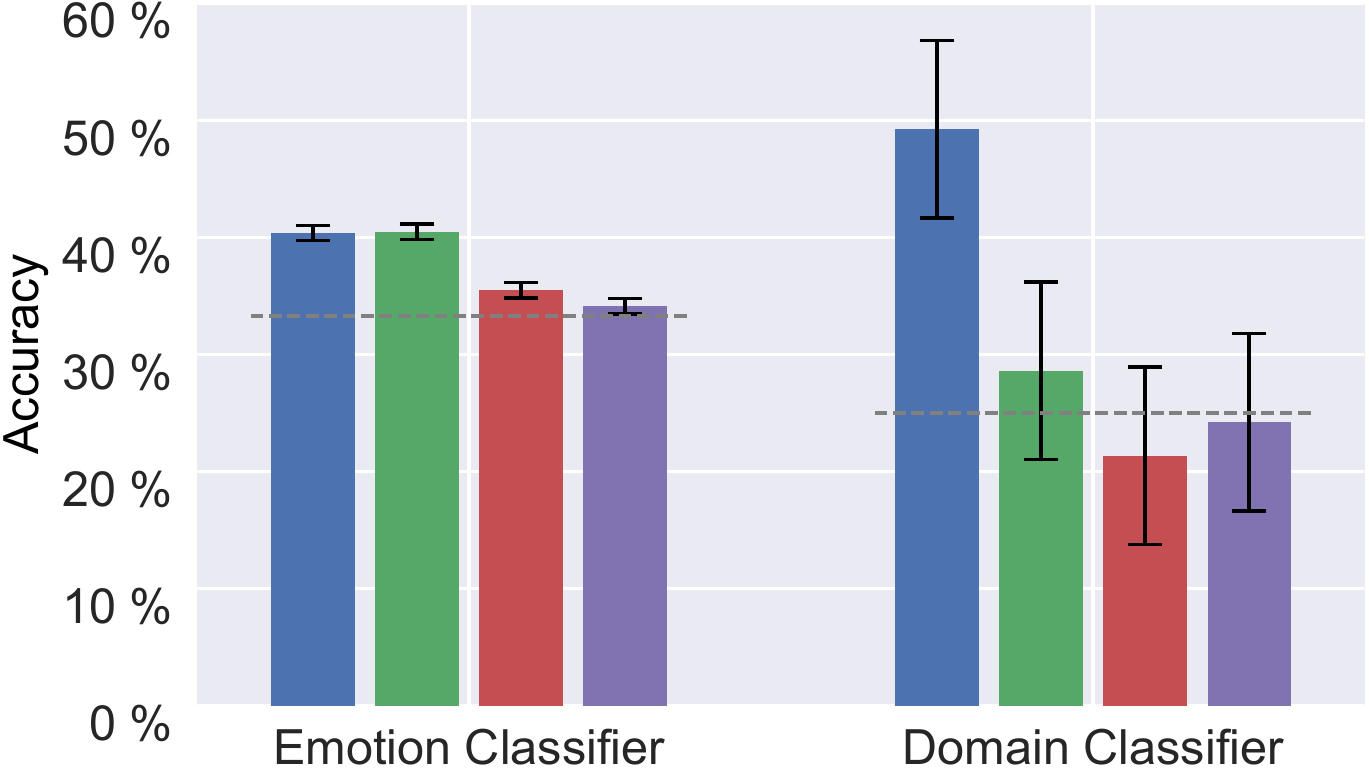}
\caption{Comparison of the mean emotion classification and data-source identification accuracies for different hyperparameters $\lambda$, averaged over 5 runs. Horizontal dashed lines represent the chance-level accuracy, and black solid lines show the empirical standard deviation.}\vspace{-.1cm}
\label{fig:adv-acc-vs-cla-acc}
\end{figure}

The classification performance of the emotion classifier depends on the choice of the hyperparameter $\lambda$, due to the revealed influence of the $d$-invariance imposing optimization scheme. Figure~\ref{fig:adv-acc-vs-cla-acc} shows final test set accuracies of the two classifier ends (emotion and dataset ID classification) of the overall architecture for different $\lambda$ choices. We consistently observe that the accuracy of the emotion classifier is not significantly impacted with increasing $\lambda$, however then starting to decrease due to high adversarial censoring leading to loss of emotion-relevant discriminative information in the latent representations. Regarding the accuracy of the domain (data-source ID) classifier, censoring accordingly with $\lambda>0$ leads to the data-source no longer be meaningfully decoded by the adversary network, while there was an observed $>$50\% data-source ID classification accuracy by the domain classifier for $\lambda=0$ baseline models, \textit{i.e.}, regular CNNs.

\subsection{EEG Classification Results}

Our method works on a very restricted dataset as described in Sec.~\ref{sec:expstudy} to test representation transfer capabilities across four emotion recognition datasets. Since in our experiments we utilize intersecting subsets of channels in each data-source and filter accordingly, as well as discard observations from specific classes for stratified sampling, the actual emotion classification task becomes highly challenging.
% It hence cannot be compared to existing EEG classification approaches learning from the original EEG dataset.

Table~\ref{tab:results-overview} shows averaged accuracies for the adversarially learned model, as well as the baseline global model. Our models achieve an above-chance classification performance for emotion recognition across all four datasets. We further showed that invariant models can be learned by reducing the leakage and maintaining a similar emotional classification quality to $\lambda=0$ cases (cf. Figure~\ref{fig:adv-acc-vs-cla-acc}).

\section{Discussion \& Conclusion}
\label{sec:discussion}

In this paper we explore robustly transferable patterns across multiple EEG emotion recognition data-sources.
We present an adversarial learning framework to unify different EEG data-sources and labels for multi-source transfer learning by finding data-source-invariant shareable information for multiple EEG-related tasks.
Our approach makes significant pre-processing steps to unify the data basis for multi-source transfer learning. Thereby, the results indicate that the pre-processing comes at the cost of classifier performance overall. However our adversarial censoring approach achieves the same classification performance as simply pooling the data domains together (\textit{i.e.}, training regular CNNs with pooled datasets with $\lambda = 0$) while giving us the opportunity to restrict the representation to be highly data-invariant ($35\%$ leakage). 
% To encourage research in this area, we publish the source code of our approach for further analysis by the research community\footnote{\url{https://github.com/philipph77/ACSE-Framework}}.
Our implementations are publicly available at: \url{https://github.com/philipph77/ACSE-Framework}.
%\footnote{\url{https://anonymous.4open.science/r/ACSE-Framework/}}.

Our work can be extended by adapting the encoder framework to be able to use different EEG input shapes according to the specified data-source, and as a result, different number of channels and sampling frequencies can be learned. We envision an adversarial shared-private model similar to~\cite{liu2017adversarial} where some channels are shared among data-sources (as in our approach) but private (data-source-specific) input can be incorporated. Our approach can also easily be adapted to learn representations that are invariant corresponding to other EEG variation factors \textit{e.g.}, participant ID, by adding an additional adversarial classifier \cite{han2020disentangled,han2021universal}.

\begin{table}[ht]
    \begin{center}
    \centering
    \caption{Mean emotion classification accuracy by the adversarially learned model and a baseline global model trained without adversarial censoring, averaged over 5 runs.} \vskip.5em
    \label{tab:results-overview}
    \resizebox{0.48\textwidth}{!}{%
    \begin{tabular}{llll}
    \toprule
    \multirow{2}{*}{\parbox{0.6cm}{}} & \multirow{2}{*}{\parbox{2.6cm}{Time-Series DNN}} &  \multirow{2}{*}{\parbox{2.3cm}{PSD Features MLP}} &   \multirow{2}{*}{\parbox{2.6cm}{Time-Series DNN \\ (binary)}} \\ \\
    \midrule
    % Local Model & $42.4\%$ & $42.7\%$ & $58.5\%$  \\
    \textbf{Global} & $40.37\% (\pm 0.65\%)$ & $40.26\% (\pm 0.36\%)$ & $57.63\%  (\pm 0.77\%)$ \\
    \textbf{Adversarial} & $40.48\% (\pm 0.70\%)$ & $38.74\% (\pm 0.65\%)$ & $58.17\% (\pm 1.63\%)$ \\
    \bottomrule
    \end{tabular}
    }%
    \end{center}
\end{table}

%\acks{Acknowledgements go here.}

\bibliographystyle{IEEEtran}
\bibliography{root}

\end{document}